# Multi-Objective Reinforcement Learning for Automated Resilient Cyber Defence

Ross O'Driscoll

*Roke Manor Research Ltd.*
Woking, UK.
Ross.ODriscoll@Roke.co.uk

Claudia Hagen

*Roke Manor Research Ltd.*
Woking, UK.

Joe Bater

*Roke Manor Research Ltd.*
Woking, UK.

James Adams

*School of Mathematics and Physics,*
*University of Surrey*
Guildford,
GU2 7XH,
UK
James.Adams@surrey.ac.uk
https://orcid.org/0000-0002-5299-6378

*Abstract*—Cyber-attacks pose a security threat to military command and control networks, Intelligence, Surveillance, and Reconnaissance (ISR) systems, and civilian critical national infrastructure. The use of artificial intelligence and autonomous agents in these attacks increases the scale, range, and complexity of this threat and the subsequent disruption they cause. Autonomous Cyber Defence (ACD) agents aim to mitigate this threat by responding at machine speed and at the scale required to address the problem. Additionally, they reduce the burden on the limited number of human cyber experts available to respond to an attack. Sequential decision-making algorithms such as Deep Reinforcement Learning (RL) provide a promising route to create ACD agents. These algorithms focus on a single objective such as minimizing the intrusion of red agents on the network, by using a handcrafted weighted sum of rewards. This approach removes the ability to adapt the model during inference, and fails to address the many competing objectives present when operating and protecting these networks. Conflicting objectives, such as restoring a machine from a back-up image, must be carefully balanced with the cost of associated down-time, or the disruption to network traffic or services that might result. Instead of pursing a Single-Objective RL (SORL) approach, here we present a simple example of a multi-objective network defence game that requires consideration of both defending the network against red-agents and maintaining critical functionality of green-agents. Two Multi-Objective Reinforcement Learning (MORL) algorithms, namely Multi-Objective Proximal Policy Optimization (MOPPO), and Pareto-Conditioned Networks (PCN), are used to create two trained ACD agents whose performance is compared on our Multi-Objective Cyber Defence game. The benefits and limitations of MORL ACD agents in comparison to SORL ACD agents are discussed based on the investigations of this game.

## I. Introduction

Cyber-attacks present a challenge to the computing networks of critical civilian infrastructure, government, and military, particularly when aided by artificial intelligence (AI) that can increase the scale and effectiveness of these attacks. To mitigate this threat many National Strategy Papers support the development of Automated Cyber Defence (ACD) [1] (in press). In the military domain a reference architecture has been created to describe Autonomous Intelligent Cyber-defence Agents (AICA) [2]. These agents could be deployed to make the operation of cyber-physical systems, such as a cargo truck, more resilient [3]. The agent seeks to maintain the critical networked systems such as truck steering, acceleration, braking and navigation so that the truck's cargo can be delivered. Any actions that the agent takes to protect the network should be carried out in a stealthy way, as the agent itself is a natural target for an attacker.

Network defence games such as the CAGE challenges provide a simulated environment to train and test blue (defensive)-agents [4], and find an appropriate response to red (offensive)-agent attacks through taking defensive actions such as restoring a machine to factory settings, or removing malware from a host. Typically the blue-agent's reward is a weighed sum of penalties for machines being compromised, or red-agent access to a sensitive area of the network [5]. This weighted sum reduces a multi-objective (MO) problem to a single objective problem. The resulting simplification introduces limitations into the model, as it fixes the trade-off, or desired utility, *a priori*, and does not provide a means for the end user to change it. The onus is on the creator of the reward function to select the scaling between the objectives in the final reward function. In the CAGE challenge example, a blue-agent could be trained to prioritise the defence of a server, but this could neglect critical user access to the server through overzealous *restore* actions. Additionally, there might be a non-linear relationship between objectives, so the agent learns a suboptimal policy. The CAGE challenges feature a *green-agent*, that simulates user activity on the network [6]. Monitoring the green agent allows the impact of the red- and blue-agent actions on users to be assessed.

This work investigates the creation of MO ACD agents, whose policy can be altered at inference time, to take into account the relative importance of the objectives and select appropriate actions as a result. The main contributions are as follows.

- A MO cyber defence game is created using the CybORG gym [4], that has both green agent objectives to maintain network functionality (resilience) and blue agent objectives to defend the network from attack.
- Two Multi-Objective Reinforcement Learning (MORL) agents are trained with online access to this gym – Multi-Objective Proximal Policy Optimisation (MOPPO) and Pareto Conditioned Networks (PCN).

- The performance of these two MORL techniques is compared by examining rollouts on the MO ACD game trading-off green agent functionality with network defence, and assessing the algorithms' ability to prioritise objectives.

## II. BACKGROUND

The network ACD problem of interest here can be described as a Markov Decision Process (MDP), with a state space $\mathcal{S}$ describing the possible states of the environment, and an action space $\mathcal{A}$ describing the possible actions. The reward function for taking a particular action $a \in \mathcal{A}$, with the system in state $s \in \mathcal{S}$ is given by $r(s, a)$. This scalar reward can be used as a learning signal for training decision making algorithms in ACD. Reinforcement Learning (RL) algorithms learn a policy $\pi$ that outputs actions that maximize the discounted reward signal during training, and have proven particularly successful in training agents for ACD [6].

### A. Multi-Objective RL

A natural extension of single-objective RL (SORL) is to use a vector-valued reward function of each state and action, with each element containing a reward for each objective. The resulting Multi-Objective Markov Decision Process (MOMDP) forms the theoretical basis for MORL [7]. For SORL the policy space can be ordered according to the expected cumulative discounted reward for each policy. By comparison MORL policies cannot be ordered in the same way as their return is a higher dimensional quantity with reward components for each objective. The optimal policy depends on the end-users priorities for one objective over another, or external dynamic costs associated with a particular policy such as the selling price of an ore, in the case of the minecart problem described in [8].

A set of potentially optimal policies can be computed, and is known as the *Pareto Front* (PF) (illustrated in Figure 1). At any point on the PF, there is no policy that is better in all objectives. MORL aims to find policies that approximate the PF as closely as possible [9].

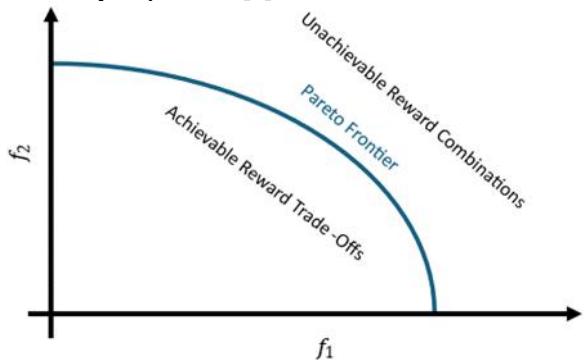

**Figure 1 – An illustration of a Pareto Front for a game with two objectives, and associated cumulative rewards $f_1$ and $f_2$. Policies inside the Pareto Front are achievable, but are inferior to those on the Pareto Front.**

Many MORL algorithms have been proposed which can be applied to MO decision making depending on factors such as the desired solution and whether the utility (i.e. the trade-off) of the problem is fixed at training time or selected during inference. There are several classes of MORL algorithms that are potentially relevant to ACD, and they are briefly summarized here.

#### 1) Single Policy Multi-Objective

The utility function that describes the end user's preferences is known in advance, so only a single policy needs to be trained. It will lie on the PF, ensuring that it is not dominated in all objectives by another policy. Like SORL, only a single policy is generated, however the rewards are handled separately depending on the algorithm, allowing guarantees on optimality that are not possible in SORL. Examples include MO Q-learning [10], which extends single objective Q-learning by extending the Q-table to encode a separate value for each objective, and MOPPO [11], which can be set up to calculate separate advantages for each objective.

#### 2) Multi-Policy Multi-Objective

If the end user's preferences are not known in advance, then multi-policy MORL can be used to generate the segments of the PF in parallel. This might be by looping over single policy algorithms with objective weightings to build up the PF, or using methods than can evaluate multiple policies in parallel to determine the PF. It is important to generate a representative and well distributed set of policies approximating the optimal PF. The appropriate policy can be selected after the user's preferences are expressed.

Multi-policy MORL techniques use additional metrics during training to evaluate if a proposed policy is dominated by the current best estimate of a PF [7]. For example, Pareto Q-Learning [12] trains multiple policies in parallel, and prunes those which are dominated by other policies. PCNs [13] maximize the hypervolume of all dominated policies between the PF and a reference point. Envelope Q-Learning [14] defines a distance between MO Q-value functions to find the PF. At inference, these models are prompted with desired preference to generate the trajectory (e.g. the target returns for each objective in the case of PCN).

#### 3) Lexicographic Algorithms

Instead of specifying the utility function to quantify the value of one objective with respect to another, lexicographic techniques specify an ordering of the multiple objectives. This categorical expression of the preferences between objectives is a fundamentally different approach to MORL. Ref. [15] demonstrates Lexicographic Optimization for MORL where objectives are fulfilled hierarchically, ensuring that higher priority objectives are optimized, at the expense of the lower order objectives. This requires an *a priori* understanding of priorities and aims to reach a lexicographically optimal rather than a Pareto optimal solution. Lexicographic RL is close in concept to the field of Safe RL where a single objective is optimized subject to a constraint.

### B. Automated Cyber Operation Gyms

Training ACD agents is typically carried out in simplified, simulated environments to speed up their training and evaluation, before moving them to more realistic environments for further fine-tuning and testing prior to deployment. The realism of the gym environment enables more rapid transfer between the simulation and the real world. The cyber gyms considered for experimentation in this paper included Yawning Titan [16] [1], CybORG [4], PrimAITE[2] and CyberBattleSim[3]. A detailed comparison of these environments has been carried out in [1]. The CybORG gym was chosen for experimentation here as it is proven in a series of open CAGE Challenges[4]. The scenarios from the challenges can be adapted into multi-objective games by decomposing the blue-agent rewards in the available scenarios into multiple objectives, and there are existing rules-based red agents for experimentation.

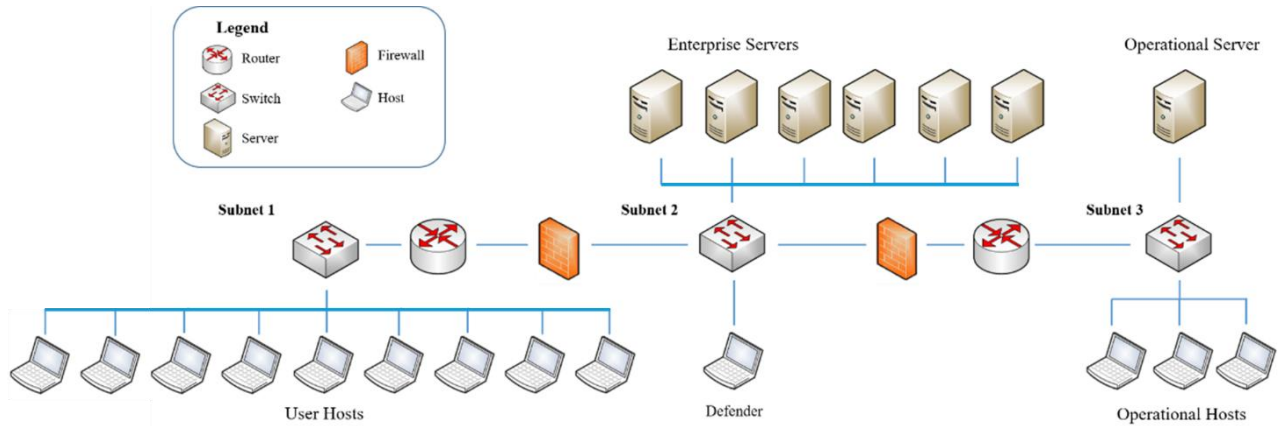

**Figure 2 – Network topology for the cyber defence game investigated here. This is a modification of the CAGE Challenge 2[4] altered by increasing the numbers of: User hosts from 5 to 9; and Enterprise servers from 3 to 6.**

*C. Related Work*

A range of different RL algorithms have been successfully applied to control problems including Monte-Carlo Tree search [17], proximal policy optimisation (PPO) [18], Deep Q-Networks (DQN) [19] and more recently Decision Transformers [20] and other large language model (LLM) approaches [21]. The appropriate algorithm depends on the available data, access to the training environment, or knowledge of the shape of a reward function [22].

RL has been used to create agents for simulated network defence games, and explored in the literature using the CybORG environment in response to the CAGE challenge [23], and further developed using a hierarchical architecture to select defensive agents suited to particular red agent strategies [24]. The effectiveness of RL agents has been investigated using the Yawning Titan environment by Acuto and Maskell [25]. They find PPO, Advantage Actor Critic (A2C) and DQN perform similarly in preventing the infection of a network for a small number of nodes. However, as the network increases in size, DQN tends to perform best with A2C performing the worst.

The importance of reward shape in ACD games is studied in [5], in particular the RL agent training is shown to be sensitive to the relative magnitude and size of the components of rewards. This indicates that training MORL agents on a decomposed reward should demonstrate different priorities and behaviour.

There is limited literature on training multi-objective algorithms for ACD. There are examples of creating red agents for penetration testing with MORL. Yang and Liu [26] consider the two objectives
1) Compromising host machines with minimal number of actions and
2) Compromising more machines on the network.
They applied multi-objective Q-learning to penetration testing in the CybORG environment [4]. They compare non-linear (Chebyshev decomposition) with linear scaling and find that linear scaling can only recover a convex optimal set of policies. They argue that the optimal set of policies is a more complex surface as it is correlated with the topology of the network. They find that some solutions are sub-optimal when using simple linear scaling and the Chebyshev Decomposition allows dominant policies to be achieved.

These works do not address the network resilience, such as maintaining critical services when subject to cyber-attack, that is crucial here.

### III. METHOD

The network layout from CAGE Challenge scenario 2 was chosen as a starting point for a multi-objective cyber defence game. This layout is shown in Figure 2. In this scenario there are two rules-based red agents: a 'meandering' agent which explores the network using a random walk, and a 'b-line' agent which simulates insider knowledge and follows a known path through the network to the operational server. The meandering agent was selected for this work.

---

[1] Python Implementation: https://github.com/dstl/YAWNING-TITAN
[2] Python Implementation: https://github.com/Autonomous-Resilient-Cyber-Defence/PrimAITE/
[3] Python Implementation: https://github.com/microsoft/cyberbattlesim
[4] Competition Repository: https://github.com/cage-challenge

### A. Multi-Objective Cyber Defence Game

Table 1, shows the reward components for the three different game set-ups used in this project. Using these components, several games were constructed for experimentation.

- Game A: Single-objective game summing the components 1, 2 and 3 used for the original CAGE challenge.
- Game B: Two-objective game trading off red agent access (component 1) with disruption to network (summing components 2 and 3).
- Game C: Two-objective cyber game trading off network security (summing components 1, 2 and 3) with *sign of life* objective (component 4).

**Table 1: Network defence game reward components.**

| Component | Reward |
|---|---|
| 1. Red access to: | |
|    User Host | -0.1 |
|    Enterprise Server | -1 |
|    Operational Host | -0.1 |
|    Operational Server | -1 |
| 2. Red disrupts operational server | -10 |
| 3. Blue agent reimages a machine | -1 |
| 4. Green agent access to port on random machine. (sign of life) | 1 |

The sign of life objective represents the impact on users of the network (green agents). It is constructed by computing the number of ports that a green agent can access on a randomly selected machines in a chosen subgroup on network, and as such it is an intrinsically noisy metric. This is motivated by the unpredictable nature of user traffic on a network. During an episode the blue-agent can start up additional services and open corresponding ports to access them. If the blue-agent restores a machine from a back-up (to remove red-agent infiltration), then these services will be stopped, and the associated ports closed. This defensive action reduces the services available to the green agent.

Following some preliminary experimentation, it was found that the blue agent maximized the second objective in Game B every time. The network from CAGE challenge scenario 2 was therefore modified to make a more challenging environment for the blue agent. The number of User Hosts was increased from 5 to 9 and Enterprise servers from 3 to 6 (see Figure 2). This increased both the action space and the observation space of the blue agent. See Appendix A for implementation details.

### B. Algorithms

Representative algorithms were identified from the literature, together with baseline implementations [27] to highlight the state-of-the-art for both single policy and multi-policy algorithms.

#### 1) Single Policy – Multi-Objective PPO

PPO algorithms [18] are a family of policy gradient methods which restrict updates to a trust region around the current policy, typically formulated as actor-critic architectures. They are robust and used in a variety of applications. They can be extended to multi-objective PPO (MOPPO) [11], by treating the reward and critic's advantage as a vector rather than a scalar. The critic's advantage is then weighted using user supplied weighting to a scalar before the loss calculation. When a single objective is used it reduces to standard PPO. A linear weighting was used in Experiments 1&2 (see Section IV), with a Chebyshev weighting also investigated in Experiment 2 (see Section IV.B.2)). As the weighting is selected *a priori*, the utility is fixed during training, and a single policy is generated per training run.

#### 2) Multi Policy - Pareto Conditioned Networks

A PCN uses a single network to estimate policies along an estimated PF [13]. During training the final reward is used as an input to the network along with the observation, so the network learns to condition output based on the expected reward. The quality of the estimated front is evaluated using several metrics, including the hypervolume between a reference point and the proposed PF, and the distribution of the policies. Once trained, PCNs can be prompted with a target reward for each objective and a time horizon for the reward, meaning the utility can be selected *a posteriori*.

### C. Model Training and Hyperparameters

The hyperparameters used in model training are shown in Table 2. Model training was carried out using an AMD Ryzen Threadripper 3970X 32-Core Processor running at 2200MHz with a Tesla V100 PCIe 32GB GPU. Typically, this took 4 hours per policy training episode in MOPPO, and 48 hours to train a PCN model.

**Table 2: Model hyperparameters used in training**

| Hyperparameter | Value |
|---|---|
| Environment: | |
|    Episode Length | 512 |
| MOPPO: | |
|    Training Timesteps | $7.5 \times 10^5$ |
|    Update Epochs | 10 |
|    Learning Rate | $2.5 \times 10^{-4}$ |
| PCN: | |
|    Reference Point | [-500.0, 6500.0] |
|    Training Timesteps | $10^7$ |
|    Batch Size | 256 |

## IV. RESULTS

Three experiments were conducted using Game A, B and C to evaluate MORL compared to SORL, and compare Multi-Policy to Single-Policy MORL algorithms.

### A. Experiment 1 – SORL vs MORL

Single objective PPO was trained on Game A. For a fair comparison the MOPPO algorithm with a single objective, and the same hyperparameters was used. This was used as a baseline to compare with MOPPO trained on Game B. To compare the multi-objective returns with the single-objective game, an equal weighting was used for both objectives in Game B in MOPPO training. The mean scalarised return evaluated during training for a typical training run is shown in Figure 3. This figure shows that MOPPO achieves a higher reward compared to PPO. Additionally, it achieves its higher return more quickly than the single objective comparator.

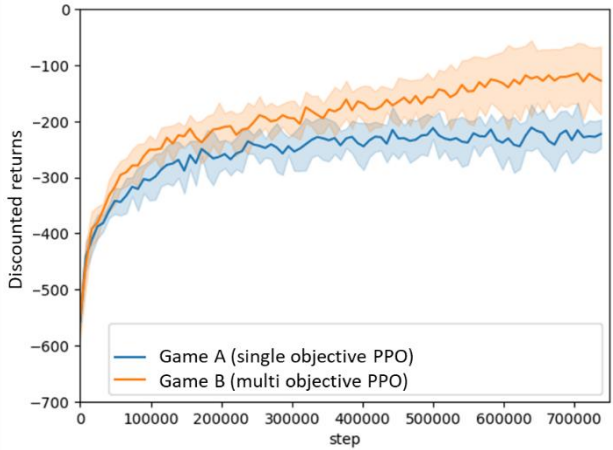

**Figure 3 – Mean discounted return of 10 training runs for single objective PPO on Game A (blue line) and MOPPO on Game B with an equally weighted sum of returns (orange line). Shaded error shows 1 standard deviation**

To compare the dynamics of the training of MOPPO against single-objective PPO, Figure 4 shows the average return per objective for each algorithm as separate plots. For Game A the components of the return are separated out into the two components. This illustrates that the "red disruption" objective (2) and (3) reaches its final value earlier in the training than the "red access" objective (1).

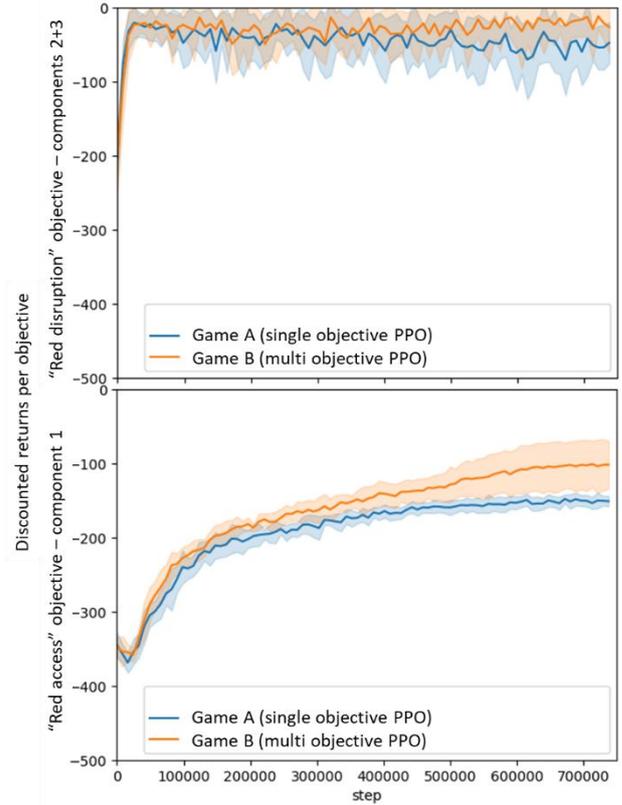

**Figure 4 – Mean discounted reward components of 10 training runs for single-objective PPO trained on Game A, and MOPPO trained on Game B with an equally weighted sum of returns. Shaded error shows 1 standard deviation.**

### B. Experiment 2 – Single Policy MOPPO

A set of policies were trained for Game C using MOPPO, to approximate the PF. A diverse range of policies were investigated by stepping through the weighting of the two policies with weights [0,1] in steps of 0.1 for each objective. The weightings referred to are

(Blue Objective: network defence, Green Objective: sign of life)

#### 1) Linear Scaling

The final policies for each linear weighted combination were evaluated for 1000 episodes. Figure 5 shows the mean performance for each point, together with error bars of one standard deviation. The error bars are larger for component (4) -- the Green Objective sign of life, partly due to the stochastic nature of this objective. As the weighting is increased towards the Green Objective, its associated mean return increases at the expense of the Blue Objective.

When the weighting placed on the Blue Objective of network defence is emphasised, the algorithm achieves similar values to those found in the single-objective Game A, Experiment 1 Figure 3. There are a cluster of policies (labelled α in the figure) with Green Objective weights [0, 0.2] which achieve similar performance and with small standard deviations. As the Green Objective weighting is increased, there is an approximately linear trade-off between the two objectives with significant increases in Green agent's access achieved, with minimal impact on Red agent ingress (Green Objective weights [0.3, 0.6] up to point β in the figure). With further increase in the Green Objective weighting there is a significant reduction in the reward due to Red agent ingress with minimal increase in the reward for Green agent's access.

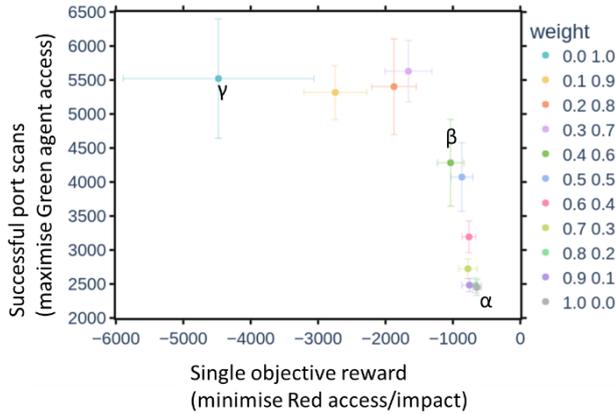

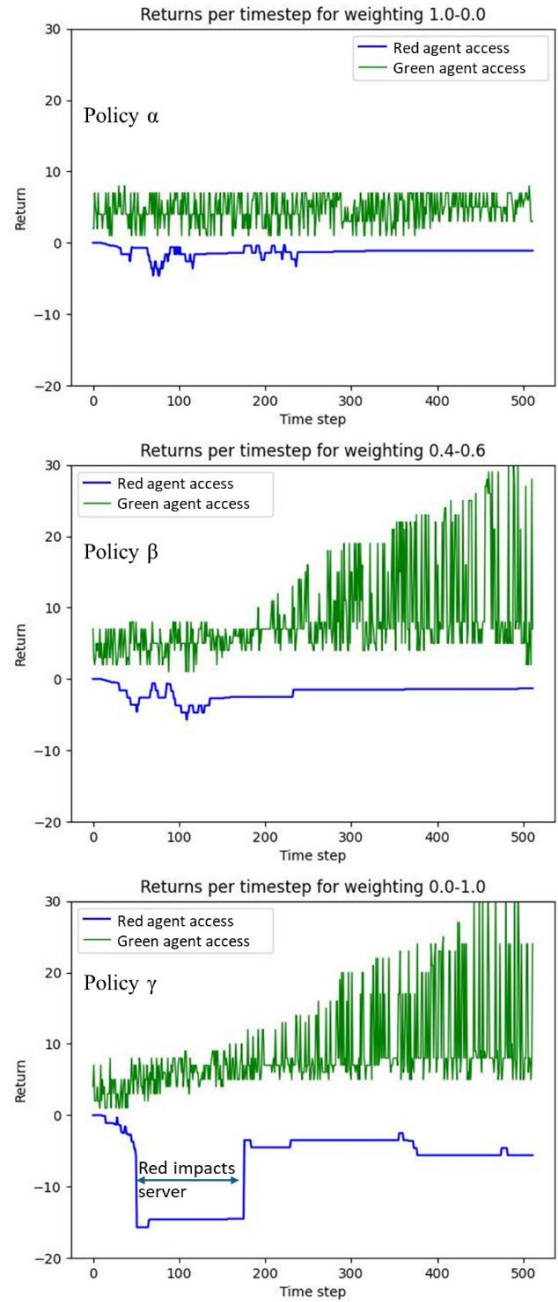

**Figure 5 – Performance space of the MOPPO algorithm trained on Game C. This should approximate the Pareto Front illustrated in Figure 1. The points step through a linear weighting between the two objectives in increments of 0.1. Each point shows the mean reward for 1000 evaluations of a policy with one standard deviation error bars.**

The *a priori* selection of regularly spaced weights gives no guarantee on how the policies are distributed along the PF, however in this case the linear spacing has resulted in a good selection of policies which are well distributed with meaningfully different capabilities.

To evaluate the performance qualitatively and illustrate the difference between the agent performance in the two objectives, runs were generated from three policy weightings, shown in Figure 5 labelled $\alpha$, $\beta$ and $\gamma$. Rollouts illustrating each of these policies are shown in Figure 6. The extreme ($\alpha$) which prioritises the single objective Game A reward (1, 0), a policy ($\beta$) which significantly increases green agent access with minimal impact on red access return (0.4, 0.6) and an extreme ($\gamma$) which maximises Green Agent access (0,1) with a catastrophic impact (labelled in figure), on the Red agent ingress due to Red agent running a command on the operational server (component 2 of the reward). This figure shows a qualitative difference between each of the policies, showing that the MOPPO algorithm can switch between prioritising the two objectives.

**Figure 6 – Example rollouts for three different policy weightings in Figure 5 (α, β and γ). Top: 1,0, middle: 0.4,0.6 and bottom: 0,1. Blue is the single objective reward, green is successful port scans (x-axis and y-axis respectively in Figure 5). Higher return is better. Note the impact on returns as the weightings change.**

An illustration of the dynamic possibilities offered with MORL is shown in Figure 7, where the defensive agent policy is switched part way through the rollout. The change in the prioritisation of the objectives results in different action choices of the autonomous agent, and a qualitatively different return for the agent metrics.

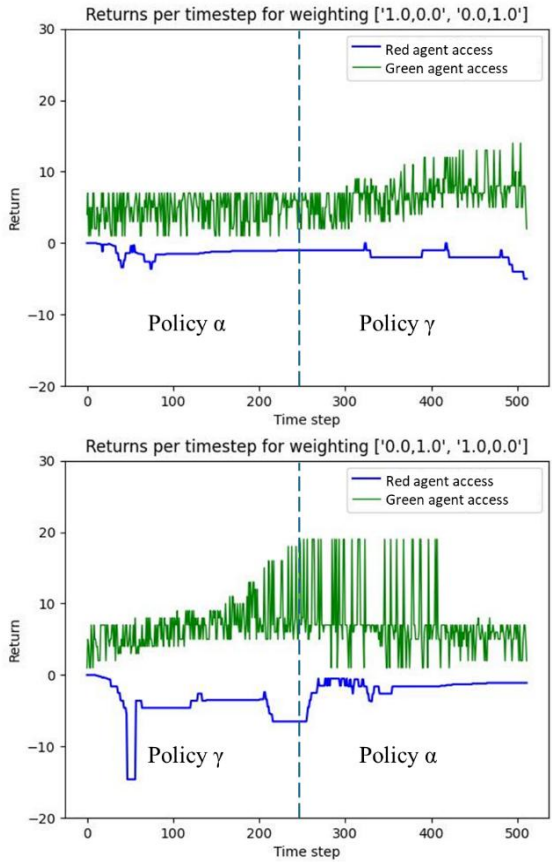

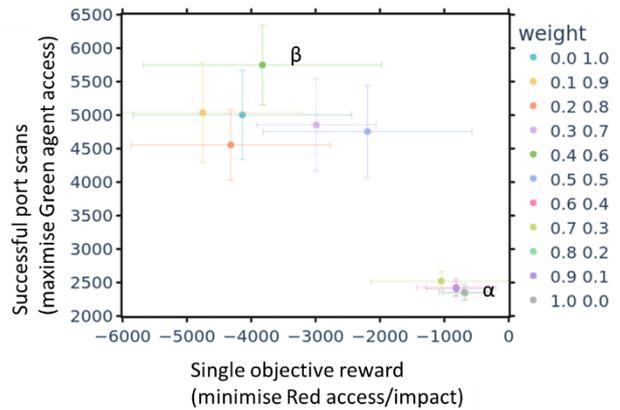

**Figure 7 – Example rollouts when switching between policies α and γ from Figure 5 at timestep 256. Top: α then γ, bottom: γ then α. Blue is the single objective reward, green is successful port scans (x-axis and y-axis respectively in Figure 5). Higher return is better. Note the impact on returns as the weightings change.**

*2) Chebyshev Scaling*

Chebyshev scalarization was trialled in this experiment, as detailed in [26] for the advantage calculation for the critic. This scalarisation has advantages for computing non-convex PFs [28], and is provably superior but no more computationally expensive than the linear scalarisation [29]. The final policies for each weighting are shown in Figure 8. There is more clustering of policies at the extremes (labelled α in the figure) and some policies, β, achieve marginally higher returns when compared with linear scaling (Figure 5).

**Figure 8 – Pareto Front computed using MOPPO with non-linear scaling (Chebyshev), to be contrasted with the linear scaling reported in Figure 5.**

*C. Experiment 3 – Multi-Policy Pareto Conditioned Networks*

A PCN model was trained on Game C, to find the PF that balances reward component (4) with the sum of reward components (1,2,3). This algorithm can be started from an initial estimate of the PF. Training experiments with no initialization, and with the PF determined from the MOPPO in Experiment 2 were carried out. At each iteration, the PCN was evaluated at several points along the current best PF estimate. The algorithm selects the resulting Pareto dominant policies as the new best PF estimate. The evolution of a training run is shown in Figure 9.

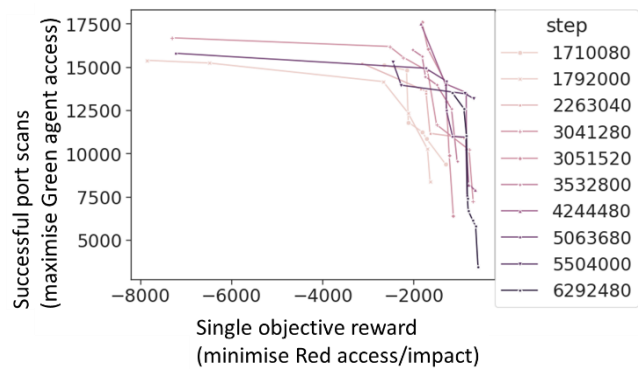

**Figure 9 - Evolution of the PCN Pareto Front during training with non-dominating policies pruned. Training time step is the colour.**

After about 2 million time steps, there is little change in the position of the PF, however it changes shape considerably. This is likely due to the way PCN selects the Pareto dominant policies when used with a highly stochastic environment, as we have here. PCN cannot distinguish between a high reward due to luck or its actions.

Prompting at inference requires specifying a time horizon, and a *desired return* (target) for the two objectives. The agent-learned policies returned similar mean values to the MOPPO, within the error bars.

To evaluate the performance, the final model was prompted at 11 points using the mean values from the 11 single policies trained in Experiment 2. These desired rewards and the resulting mean and standard deviation of returns of the rollouts are shown in Figure 10.

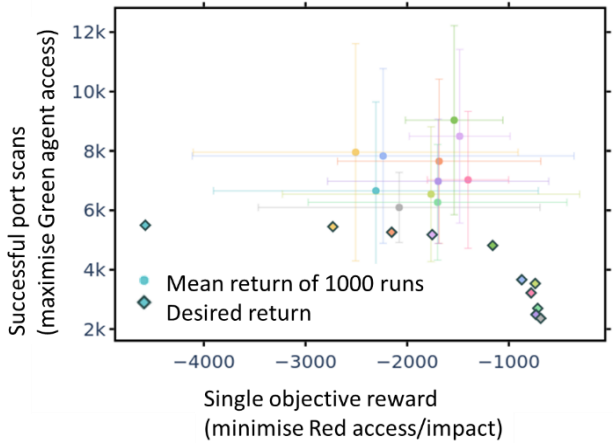

Figure 10 – Evaluation of PCN model after training for 10 million timesteps. The desired, target reward prompts are shown along with the resulting mean and standard deviation of 1000 evaluations. Note PCN uses full target returns rather than discounted.

In most cases, the Green Agent objective (y-axis) returns exceeded those of MOPPO, in some cases significantly. The returns for minimizing red access (x-axis) were generally lower than the MOPPO results with occasional larger values. The standard deviation of the PCN evaluations is very large indicating that the policies are not reliable and, unlike MOPPO, there is no clear trade-off between objectives in the evaluation results – i.e. prompting the model typically results in the same returns for the objectives.

V. DISCUSSION

The results described in the previous section indicate that there are some benefits from using MORL algorithms. In the case of the single-policy experiments with MOPPO, Experiment 1 indicates that MORL shows both faster training, at approximately 30% faster for the objective shown in Figure 3 and approximately 17% higher terminal rewards compared to SORL. This might be because one objective was significantly easier to learn than the other or breaking down the objectives allows some aspects of curriculum learning to be used to improve performance. The ACD agent might also learn to prevent red agent access to the operational server and learn strategies that do not require reimaging of machines. The red agent will always have a foothold in the system so there will always be red access to some machines.

Experiment 2 demonstrates that the MOPPO algorithm can be used to determine the PF for the two-objective game. Individually training policies from weightings across the PF is computationally expensive, and does not scale well to higher dimensions, where more objectives would rapidly increase the cost. The qualitative difference between strategies illustrates that a change in policy on-the-fly produces a meaningful change in the functionality of the network. A hierarchical controller, informed by external intelligence on likely attacks, and network user requirements could be used to select between the points on the PF to ensure the most appropriate policy is selected. The ability to tune the response of the autonomous agent might make the practical response to ongoing cyber incident, or dynamic operational requirements more fluent.

Experiment 3 demonstrated a multi-policy algorithm for the determination of the PF for Game C. The evaluation of this model indicates that its behaviour did not change significantly when prompted with different desired rewards. It should be noted that PCN models are typically poorly performing in stochastic environments. There are two sources of randomness in Game C – the random behaviour of the Red agent, and the random requirements of the Green agent on the network. This poor performance in stochastic environments is shared with offline decision transformer models [30]. Some literature has suggested alterations to PCN that improve its performance in stochastic environments [31].

Some additional experiments were carried out with Lexicographic MORL [15]. The implementation of this algorithm meant experiments were slow on Game C, and at the time of submission did not achieve as large returns as MOPPO. This algorithm might be better suited to enforcing constraints in MORL problems, as the uncertainty in the return for prioritized objectives is small.

The three experiments here indicate that single-policy MORL is the most effective at ACD, showing qualitative change in strategy as the weighting between the different objectives was changed. Further investigation of this technique would be interesting in scenarios with more objectives for the Green Agent, such as accessing particular databases, or additional objectives such as deploying honeypots. The multi-policy MORL might benefit from PCN techniques that are more robust to stochastic environments.

VI. CONCLUSION

A multi-objective ACD game was created, including two objectives relating to both network defence, and functionality of the network. MORL algorithms based on both Single-Policy and Multi-Policy techniques were evaluated on this game. It was found that the Single-Policy MOPPO algorithm could be trained both faster, and more effectively on the ACD game. A Pareto Front for this algorithm was calculated by training models with different weightings between the objectives. The models showed qualitatively different performance between network defence, and green agent access during rollouts. The multi-policy algorithm PCN was trained on the same ACD game, and a PF was computed with this model. The rollouts from this algorithm did not produce qualitatively different behaviour, perhaps because of the difficulty this algorithm had with the noisy environment, and performance metrics used in this game.

ACKNOWLEDGMENT

The authors would like to thank Alex Revell for helpful discussions about the direction of this work. Research funded by Frazer-Nash Consultancy Ltd. on behalf of the Defence Science and Technology Laboratory (Dstl) which is an executive agency of the UK Ministry of Defence providing world class expertise and delivering cutting-edge science and technology for the benefit of the nation and allies. The research supports the Autonomous Resilient Cyber Defence (ARCD) project within the Dstl Cyber Defence Enhancement programme.

APPENDIX A - MODIFICATIONS TO CYBORG 2

Game A and B were trained on the original CAGE challenge 2 topology with SO-PPO (game 0) and MOPPO (game 1) agents respectively. The results for meander and b-line red agents are shown for 100 time steps in Figure 11 for 10 training runs. Note how game B achieves near perfect scores for these runs.

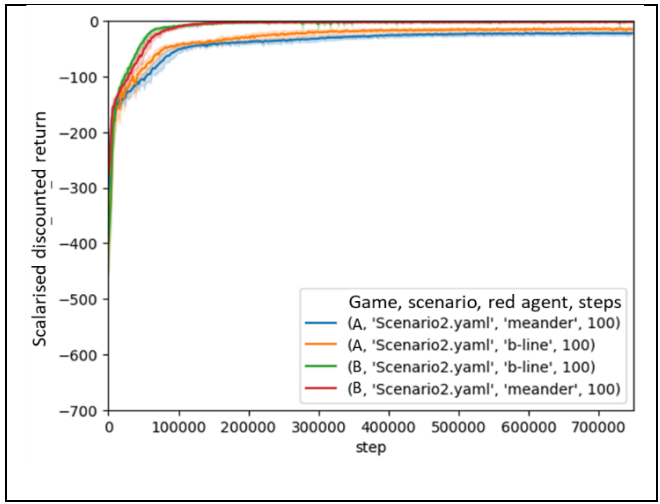

**Figure 11** – Mean discounted return of 10 training runs for single (A) and multi (B) -objective PPO. Shaded error shows 1 standard deviation. B-line vs meander red agents for CAGE Challenge2 standard set up (Scenario2) over 100-time steps.

It was decided to make the game more challenging for the blue agent against these simple adversaries. First the number of time steps was increased to 512 (Figure 12) allowing more time for the red agent to propagate through the network, secondly the topology was made more complex (Figure 13) which has the effect of increasing the observation and action space for the defending agent and increasing the complexity of the Pareto Front (Yang and Liu [26]). The specific modifications are as follows.

User subnet: User Host 4 was replicated 4 times to create User Host 5-9.

Enterprise subnet: The hosts Enterprise0, Enterprise1 and Enterprise2 were replicated to create Enterprise3, Enterprise4 and Enterprise 5 respectivel. For example, in CAGE challenge 2 only Enterprise2 can access the Operational Server, while in the modified game, Enterprise 2 and 5 can access it.

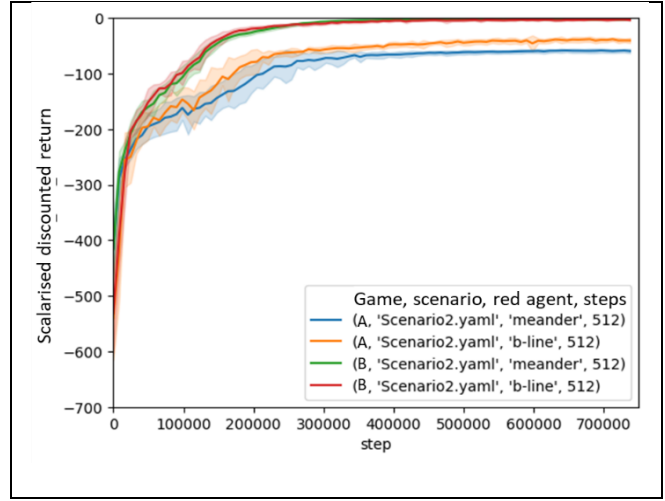

**Figure 12** – Mean discounted return of 10 training runs for single (A) and multi (B) -objective PPO. Shaded error shows 1 standard deviation. B-line vs meander red agents for CAGE Challenge2 standard set up (Scenario2) over 512-time steps.

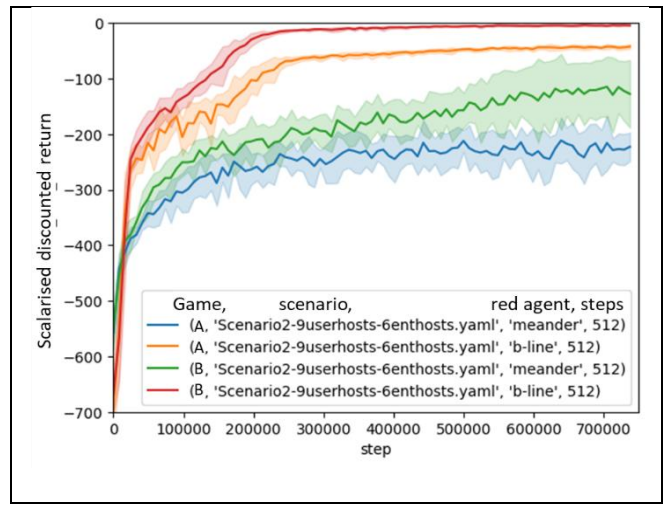

**Figure 13** – Mean discounted return of 10 training runs for single (A) and multi (B) -objective PPO. Shaded error shows 1 standard deviation. B-line vs meander red agents for this project's non-standard set up (Scenario2 with 9 user hosts and 6 enterprise hosts) over 512-time steps.

In all cases the meander agent is more challenging than the b-line agent for the blue agent to defend against. The more challenging agent better demonstrates the principles of MORL and Pareto Front recovery.

## APPENDIX B - HYPERPARAMETERS

### PPO

```
"gamma": 0.99
"weights": [1.0]
"total_timesteps": 750000
"eval_freq": 50000.0
"game": 0
"wght_strat": "linear"
"scenario":       "Scenario2-9userhosts-6enthosts.yaml"
"redagenttype": "b-line"
"num_envs": 16
"n_minibatch": 16
"num_steps": 512
"updt_epoch": 10
"learning_rate": 0.00025
"num_episodes": 20
"seed": 4
"steps_per_iteration": 100,
"anneal_lr": True,
"clip_coef": 0.2,
"ent_coef": 0.01,
"vf_coef": 0.5,
"clip_vloss": True,
"max_grad_norm": 0.5,
"norm_adv": True,
"target_kl": None,
"gae": True,
"gae_lambda": 0.95,
```

### PCN

```
"ref_point": [-500.0, 6500.0]
"max_return": [0.0, 10000.0]
"max_steps": 512
"total_timesteps": 10000000
"batch_size": 256
"game": 6
"scenario":       "Scenario2-9userhosts-6enthosts.yaml"
"redagenttype": "meander"
"known_pareto_front":"[[-4576.593262,5497.799805],[-2731.05542,5448.600098],[-2153.070068,5262.200195],[-1753.648071,5180.649902],[-1159.431396,4817.149902],[-872.6861572,3661.550049],[-780.8148193,3217.899902],[-740.6045532,3534.550049],[-716.069397,2698.5],[-733.4193115,2488.300049],[-686.8294678,2360.149902]]"
"scaling_factor": 1,1,0.1
"learning_rate": 1e-3
"num_er_episodes": 20
"max_buffer_size": 50
"num_model_updates": 50
"gamma": 1.0
"batch_size": 256
"hidden_dim": 64
"noise": 0.1
"project_name": "MORL-Baselines"
"experiment_name": "PCN"
"num_step_episodes": 10
"num_points_pf": 100
```